\documentclass[10pt,reqno]{amsart}

\usepackage{amsmath,amssymb}
\usepackage{latexsym}
\usepackage{epsfig}
\usepackage{cite}
\usepackage[english]{babel}
\usepackage{booktabs}
\usepackage{graphicx}
\usepackage{tikz}
\usepackage{pgfplots}
\usepackage{eurosym}
\usepackage[inner=4cm,outer=2cm]{geometry} 
\usepackage{amsthm}

\numberwithin{equation}{section}

\numberwithin{equation}{section}

\title{Gravitational waves, 100 years later}
\author{S.L.~Cacciatori}
\address{Universit\`a dell'Insubria, dipartimento di scienza e alta tecnologia, via Valleggio 11, 22100, Como, Italy and INFN, via Celoria 16, 20100, Milano, Italy}

\begin{document}
\maketitle

\begin{center}
\hspace{1.5cm}{\it Talk given at Accademia Lombarda di Scienze e Lettere, Palazzo Brera, Milano, Italy, \\01 March 2018} 
\end{center}

\begin{abstract}
We present a short communication on the recent direct measurements of gravitational waves, made possible by the joint efforts of the LIGO and the Virgo projects. These notes are devoted to a broad audience. In the first
part we will present the basic ideas of the theory of relativity, leading to the prediction of the existence of gravitational waves. The second part is devoted to present the main results and consequences of the discovery. 

\

\noindent
RIASSUNTO. In questa breve nota vogliamo esporre i principali risultati riguardanti la rivelazione diretta di onde gravitazionali nel periodo che va dal settembre 2015 al marzo 2018, rese possibili dagli sforzi congiunti dei 
progetti LIGO e Virgo. Data l'importanza di tali risultati, ci si vuole rivolgere ad un pubblico non necessariamente di specialisti, minimizzando le espressioni matematiche in favore di una spiegazione dettagliata dei fenomeni
fisici pur senza sacrificarne la precisione. La nota \`e dunque divisa in due parti: nella prima parte verranno esposte le idee principali che hanno portato Einstein a sviluppare la teoria della relativit\`a ristretta prima e quella generale poi, 
e come queste abbiano rivoluzionato la nostra concezione di spazio e di tempo. Nella seconda parte vedremo come la teoria di Einstein preveda l'esistenza di onde gravitazionali. Vedremo in che modo esse sono state misurate,
e quali siano le importantissime conseguenze di tale scoperta.
\end{abstract}

\section{Einstein and the new vision of the universe}

\subsection{Relativity of space and time}
In the {\it annus mirabilis} 1905, with the work {\it Zur Elektrodynamik Bewegter K\"orper} \cite{Einstein-1}, Albert Einstein started his revolutionary modification of the conceptions of space and of time, replacing them with the more fundamental concept 
of {\it spacetime}. The Einstein's spacetime, whose geometry has been clarified by Hermann Minkowski in 1908, \cite{Minkowski}, is more known under the name of Minkowski's spacetime. It is a four dimensional space, with three spatial directions and one 
temporal direction. But this does not mean that it can be obtained simply by adding to the three spatial directions a fourth one, the line of time. The Einstein's spacetime is an absolute object just like it was the space for Newton. The point is now that 
spacetime does not split in a natural way in space and time. Such a splitting can be done only after choosing a reference frame, that is by placing everywhere a clock in such a way to be sure that all clocks tick in the same manner and to be able to 
synchronise them. If all these operations can be performed, that is if all our clocks tick in the same way and remain synchronised, then we have defined a time, the timelike direction, and, thus, also the notion of space. The space is the set of points of 
spacetime that correspond to a fixed instant ``$t$''. It is the set of all synchronised clocks indicating the same time.\\
With this definition, the timelike direction, as well as the flowing of time, depends on the choice of a reference system. If a group of sailors sitting on the benches in the dock of a haven can verify that their clocks tick identically and are perfectly
reciprocally synchronised, and the same is realised by a group of sailors sitting down on the deck of a boat that is outgoing the harbour, each group will see that the clocks of the other group will appear not to be synchronised but signing different times.
But there is more: if at a certain time one of the clocks at land indicated exactly the same as one of the clocks of the boat, after a while this would be no more true and each one of the two sailors, owners of the clocks, would realise that the time for the
other sailor is flowing differently than for him, despite the clocks, let assume, are identical by construction.\\
All this happens because space and time are relative and not absolute: they depend on the choice of a reference system. The spacetime is the only thing that can be absolute without conflicting with physical principles. In any inertial frame we are
allowed to adopt any convention to define a point of space, as long as it is realisable with an experiment compatible with the laws of physics, valid in the same way in any inertial system. This means that (assuming they are inertial) each sailor in the
above example can assume to individuate an arbitrary point of the space, for example, by identifying it with the tip of the forefinger of his right hand. But, then, the sailors on the dock will see the points indicated by the sailors on the boat as changing
time by time, since the boat is moving. And vice versa. Despite both have the same right of deciding what a point of space is, they cannot be right at the same time if we assume that indeed a space does exist. The best we can do is to assume that ``the space''
does exist. After that, it could be the one defined by the sailors at land, or the one of the sailors on the boat or none of them. And, obviously, there is no way for the sailor to establish who is right or, if it is not, what the real space is. This is the problem of 
absolute space in Newtonian physics: we must introduce it, but we cannot ``see'' it.\\
In the Einstein viewpoint this problem disappears, since absolute space does not exist, and each frame, the dock and the boat, has its own space, compatibly with the physical laws. The space looses its absoluteness in favour of an absolute spacetime
thus paying the price that also the time must become relative. {\it But a new problem arises now!} In order to describe the physics and the evolution of the world around us, we need to distinguish time from space, since we have to describe the change
of a given system when time flows. So, if time flows in a different way as we change the observer, does the laws of physics also change accordingly to the observer?\\
The answer is {\it Yes} and {\it No}, but, of course, requires some explanation: {\it Yes} because we cannot avoid the changes of the explicit expression of any formula as function of space and time if space and time change; {\it No} in the sense that the
new formulas keep the same form after the change, in such a way that, when we translate them in words, they give the same law. One expresses this by saying that formulas are {\it  covariant}.
\subsection{Special Relativity}
Above we referred to the laws of physics, by these meaning the Newton's laws of mechanics. In this respect we considered inertial frames, since the concept of an inertial system is basic in Newton's mechanics. {\it Special Relativity} is a theory
of relativity based on such fundamental concept and, therefore, it privileges inertial systems.\\
{\it A reference system is said ``inertial'', if any pointlike mass it contains, which is isolated so that no forces act on it, moves with constant velocity (eventually zero) along a straight line (uniform rectilinear motion).}\\
Any uniform rectilinear motion in an inertial frame is said to be an {\it inertial motion}. \\
The covariance of the theory must then be realised into the inertial frames, which means that the laws of physic must keep the same form when we pass from an inertial frame to another one, and, therefore, they are particularly relevant when formulated in
such the frame. After that, it is quite simple to determine the transformation rules telling us how to pass from an inertial frame to another, as well as the mathematical expression of the physical laws transforming according to these rules and, finally, how
they can be written in a way that is independent from the specific choice of an inertial system. This requires the introduction of the concept of tensors, objects whose definition is independent from the choice of an inertial frame, in the same way as
the components along the coordinate axes of a vector on a plane depend on the choice of the axes, but the vector itself, of course, does not. Here, we do not intend to deduce nor discuss in details such transformation rules. Nevertheless, let us
see in a moment what are the first principles on which they are determined, and a couple of consequences that will be useful later on. The point is that, in passing from an inertial frame to another one, space and time are modified, without changing
the spacetime, so that the following rules are satisfied:
\begin{itemize}
 \item Since both reference systems are inertial, each inertial motion must be transformed into another inertial motion;
 \item In passing from an inertial frame to another, space and time are modified in such a way to keep invariant the modulus of an {\it absolute velocity}, that is, a special velocity whose modulus $V$ is independent on the choice of the (inertial)
 reference system in which the measurements are performed. 
\end{itemize}
The first issue is perfectly obvious and compatible with our common sense. In particular, it allows us to conclude that, given a particular inertial frame, every other inertial frame is made by observers all moving with exactly the same velocity
(w.r.t. the fixed frame, possibly with the axes oriented in an arbitrary way,  and with a different choice of the origin of space and time) so that they are reciprocally at rest, and appear to have an inertial motion as seen by observers at rest in the first frame
(and vice versa). On the opposite, the second issue is completely counterintuitive, very far from our common sense originated by our daily experience. If a car is shooting at 120km/h on the highway and we run after it at 100km/h, we will see it 
moving away from us at a speed of 20km/h. Or, at least, this is what it will appear to us: the speed of a car depends on the reference frame with respect to which it is moving. {\it But it is not the same for the absolute speed $V$!} This is what the
second issue is claiming. If a car is shooting at a speed $V$ and we run after it with speed $V/2$, we will not see it moving away from us with velocity $V/2$ but with velocity $V$, as if we were at rest. This may look quite odd, but it is exactly
what was postulated by Einstein, with $V$ equal to the speed of light, after an experiment by Michelson and Morley at the end of the XIX century. \\
The point is that our daily perception of the world is approximated, and the conclusions we can deduce from them may be wrong too when we check them with sufficiently high accuracy. This is why physics tries to found its principles on
objective experiments, which are experiment independent from our common sense as much as possible. 
\subsection{The Michelson and Morley experiment} 
It is interesting for us to discuss the Michelson and Morley experiment because the apparatus they invented for their experiment is substantially the same used to prove the existence of gravitational waves.\\
The most precise measurement of the speed of light performed on the Earth has given the result
\begin{align*}
 c=299792458m/s
\end{align*}
with an error on the decimals that we don't care for reasons that will be clear in the next section. We also know that the Earth is nearly $150\cdot 10^6$Km far away from the Sun, and moves around it in one year, t.i. nearly $3.15\cdot 10^7$ seconds.  
It follows that the Earth runs around the Sun with a velocity $v$ of about 
\begin{align*}
 v\approx 30Km/s=30000m/s,
\end{align*}
much larger than the precision at which we know the speed of light.\\
Suppose we want to measure the speed of the light emitted by a fixed star belonging on the plane of the terrestrial orbit. Then, there will be a period of the year when the Earth is moving toward the star with velocity $v$, and, of course, six months later it will
move in the opposite sense with the same velocity. A simple reasoning shows that if we measure the speed of the light of the fixed star in the two periods, we should expect to get the results $c_+=c+v$ and $c_-=c-v$ in the  respective periods of the year. 
In particular, we should get $c_+-c_-=2v$, sensibly observable since it is much larger than the precision in measuring $c$.\\
In order to measure such differences of velocities, Albert Abraham Michelson, an American physicist of Prussian origin, alone in 1881 \cite{Michelson-1} \cite{Michelson-2} A. A. Michelson, {\it Relative Motion of Earth and Aether,}  Philosophical Magazine 
8 (1904): 716--719; and then with the help of the American Chemist and Physicist Edward Williams Morley in 1887 \cite{MichelsonMorley-1} \cite{MichelsonMorley-2}, realized the following apparatus, {\it the Michelson and Morley interferometer}.

%

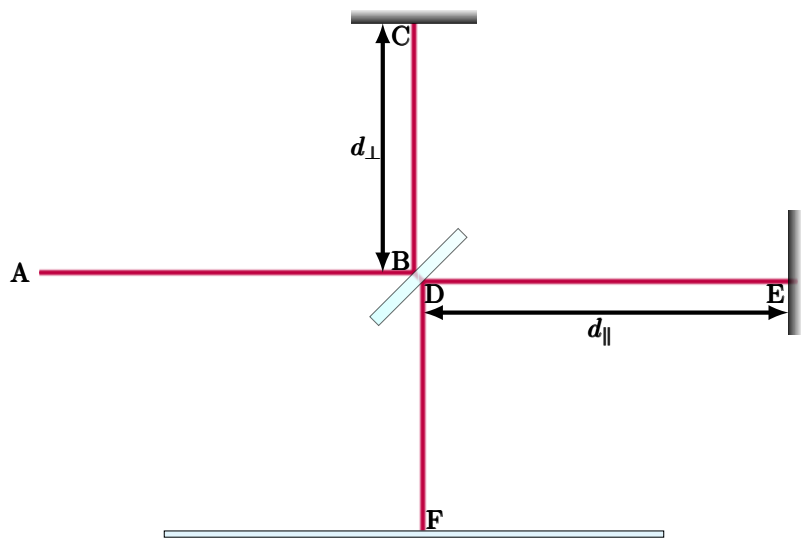
\begin{figure}[!htbp]
\begin{center}
\begin{tikzpicture}[>=latex,scale=0.83]  
\draw [purple] (-6,0) -- (0,0);
\draw [purple,line width=1pt,opacity=0.8] (-6,0) -- (0,0);
\draw [purple,line width=1.5pt,opacity=0.6] (-6,0) -- (0,0);
\draw [purple,line width=2pt,opacity=0.4] (-6,0) -- (0,0);
\draw [purple,line width=2.5pt,opacity=0.2] (-6,0) -- (0,0);
\draw [purple,line width=3pt,opacity=0.1] (-6,0) -- (0,0);
\draw [purple] (0,0) -- (0,4);
\draw [ultra thick,<->] (-0.5,0) -- (-0.5,4);
\draw [purple,line width=1pt,opacity=0.8] (0,4) -- (0,0);
\draw [purple,line width=1.5pt,opacity=0.6] (0,4) -- (0,0);
\draw [purple,line width=2pt,opacity=0.4] (0,4) -- (0,0);
\draw [purple,line width=2.5pt,opacity=0.2] (0,4) -- (0,0);
\draw [purple,line width=3pt,opacity=0.1] (0,4) -- (0,0);
\draw [purple] (0,0) -- (0.1414,-0.1414);
\draw [purple,line width=1pt,opacity=0.8] (0,0) -- (0.1414,-0.1414);
\draw [purple,line width=1.5pt,opacity=0.6] (0,0) -- (0.1414,-0.1414);
\draw [purple,line width=2pt,opacity=0.4] (0,0) -- (0.1414,-0.1414);
\draw [purple,line width=2.5pt,opacity=0.2] (0,0) -- (0.1414,-0.1414);
\draw [purple,line width=3pt,opacity=0.1] (0,0) -- (0.1414,-0.1414);
\draw [ultra thick,<->] (0.1414,-0.6414) -- (6,-0.6414);
\draw [purple] (0.1414,-0.1414) -- (6.1414,-0.1414);
\draw [purple,line width=1pt,opacity=0.8] (0.1414,-0.1414) -- (6.1414,-0.1414);
\draw [purple,line width=1.5pt,opacity=0.6] (0.1414,-0.1414) -- (6.1414,-0.1414);
\draw [purple,line width=2pt,opacity=0.4] (0.1414,-0.1414) -- (6.1414,-0.1414);
\draw [purple,line width=2.5pt,opacity=0.2] (0.1414,-0.1414) -- (6.1414,-0.1414);
\draw [purple,line width=3pt,opacity=0.1] (0.1414,-0.1414) -- (6.1414,-0.1414);
\draw [purple] (0.1414,-0.1414) -- (0.1414,-4.1414);
\draw [purple,line width=1pt,opacity=0.8] (0.1414,-0.1414) -- (0.1414,-4.1414);
\draw [purple,line width=1.5pt,opacity=0.6] (0.1414,-0.1414) -- (0.1414,-4.1414);
\draw [purple,line width=2pt,opacity=0.4] (0.1414,-0.1414) -- (0.1414,-4.1414);
\draw [purple,line width=2.5pt,opacity=0.2] (0.1414,-0.1414) -- (0.1414,-4.1414);
\draw [purple,line width=3pt,opacity=0.1] (0.1414,-0.1414) -- (0.1414,-4.1414);
\shade [left color=cyan!20!white, right color =cyan!5!white, rotate=45, opacity=0.6] (-1,0) -- (1,0) -- (1,-0.2) -- (-1,-0.2) -- cycle;
\draw [ultra thin, rotate=45,opacity=0.6] (-1,0) -- (1,0) -- (1,-0.2) -- (-1,-0.2) -- cycle;
\shade [top color=gray!20!white, bottom color=black] (-1,4) -- (1,4) -- (1,4.2) -- (-1,4.2) -- cycle;
\shade [right color=gray!20!white, left color=black] (6,-1) -- (6,1) -- (6.2,1) -- (6.2,-1) -- cycle;
\filldraw [cyan!10!white] (-4,-4.1414) -- (4,-4.1414) -- (4,-4.2414) -- (-4,-4.2414) -- cycle;
\draw [ultra thin] (-4,-4.1414) -- (4,-4.1414) -- (4,-4.2414) -- (-4,-4.2414) -- cycle;
\node at (-6.3,0) {\pmb A};
\node at (-0.2,0.2) {\pmb B};
\node at (-0.2,3.8) {\pmb C};
\node at (0.3414,-0.3414) {\pmb D};
\node at (5.8,-0.3414) {\pmb E};
\node at (0.3414,-3.9586) {\pmb F};
\node at (-0.75,2) {$\pmb {d_\perp}$ };
\node at (3,-0.95) {$\pmb {d_\parallel}$ };
\end{tikzpicture}
\caption{ The Michelson and Morley interferometer}
\end{center}
\end{figure}

\noindent It is based on a simple but efficient principle: let suppose that luminous signal comes from $A$, toward a beam splitter in $B$, that is a lens that splits the signal into two beams, the first one following the path $BCBDF$, and the second one following
the path $BDEDF$, where in $C$ and $E$ there are two mirrors and in $F$ a screen. The two beams thus follow essentially the same path, except the parts $BCB$ for the first one and $DED$ for the second one. In $F$ they will meet again, there
producing an interference figure, which will be a luminous point if they arrive with the same phases and a dark point if they arrive with the opposite phases. This difference of phases depends on the differences in the traveled paths. Suppose for example
that from $A$ it is arriving the light from the fixed star and the apparatus is moving in the $AB$ direction with velocity $v$ (the velocity of the Earth). Then, if our common sense is right, the path $DE$ is traveled by the light with velocity $c-v$ whereas
$ED$ is traveled with velocity $c+v$. On the opposite, $BC$ and $CB$ will be traveled with the same velocity. The difference between the phases will depend on these different behaviours and on the lengths $L_\parallel$ and $L_\perp$ of the two arms. \\
If we rotate the system by $90$ degrees, then the roles of the two arms will interchange and the difference of phases will change. So suppose that we first take the former configuration, like in the picture, and regulate $L_\perp$, so varying the difference 
between the phases until the interference becomes negative and the screen becomes completely dark. If next we rotate the apparatus then interchanging $L_\parallel$ with $L_\perp$, the light will now travel the arm $L_\perp$  back and forth with different 
velocities, and the other arm with the same velocity so that we expect a change in the difference of the phases and, so, a new interference picture should appear. \\
But no interference pictures appeared in the Michelson and Morley experiment, exactly as if both the arms had traveled back and forth with exactly the same velocity.\\
Several possible explanations appeared, based on different strange properties of aether, but the simplest one, proposed by Einstein, was that aether actually does not exist and the light moves in vacuum with an absolute velocity. Of course, his proposal
was a theoretical extrapolation since we are living on the Earth that does not constitute an inertial frame. However, each measurement by Michelson and Morley took a short time, during which the displacement of the Earth is very small
compared to the radius of its orbit, of about $150\cdot 10^6$Km, so that in such time intervals its motion can be considered as inertial. Nowadays we know that the postulate of absoluteness of the speed of   can be replaced by the requirement that 
the space should be isotropic and homogeneous, and that time must be homogeneous, for any choice of an inertial frame. Together with the preservation of the principle of inertia, one can show that this allows to prove the existence of an absolute velocity 
$V$, which may be infinite (in such case one gets the usual Galilean theory of relativity) or finite. In light of this, the fact that the value of $V$ coincides with the speed of light $c$ is not a principle but a mere accident. On the opposite, we could say that it is
the speed of light to be (accidentally) equal to the absolute velocity, as a consequence of the covariance of the Maxwell equations, which are the equations describing the electromagnetic phenomena. Whereas the existence of an absolute velocity
is a general kinematical fact related to fundamental principles, the Maxwell equations are just a specific model describing light, and the requirement that they maintain the same form in any inertial frame fixes the speed of light to coincide with the
absolute velocity.

\subsection{Some consequences of Special Relativity}
Once accepted the absoluteness of the speed of light, there is no more reason to improve its measurements with respect to fixed choices of units of length and of time! After defining the unit of time, the second $s$, we can define the unit of length, the meter 
$m$, so that the speed of light is exactly (and not an approximate value)
\begin{align*}
c=299792458m/s. 
\end{align*}
In this sense the velocity looks more fundamental then a length, which is now derived by the definition of time and velocity. This allows getting precise definitions of some units of measure. The Cesium is a radioactive material that, when excited, emits
a particular electromagnetic radiation oscillating 9192631770 times each second. Indeed, one uses this precise measure in order to define what a second is: it is the time employed by the radiation of Cesium to have the above number of oscillations.
With this precise definition of second, we can define the meter to be the length traveled by light in vacuum in a fraction 1/299792458 of one second.\\
But there are at least three further consequences of Einstein's postulates that we want to mention now, without pretending to deduce any of them.
\subsubsection*{Lengths contraction}
Let us consider an observer at rest just to the side of the highway of an inertial frame, looking a car of length $L_0$ running with velocity $v$. To him, the length of the car will be
\begin{align*}
 L=L_0\sqrt {1-\frac {v^2}{c^2}} 
\end{align*}
so that it will appear shorter than with respect to the driver (whereas he will agree about the other sizes, transversal to the direction of motion). This phenomenon is called {\it contraction of the length}.
\subsubsection*{Time dilation}
If the above observer measures the lapse of time between two successive cardiac pulses of the driver, that for the pilot himself is $\Delta t_0$, he will find
\begin{align*}
 \Delta t=\frac {\Delta t_0}{\sqrt {1-\frac {v^2}{c^2}}},
\end{align*}
which is a longer time. This is the {\it time dilation}. All the processes in the cabin of the car will appear to be slowed down from the observer's viewpoint, standing by the road.
\subsubsection*{Rest energy} 
One of the most famous results of the Einstein's theory of special relativity is that the mass can be converted into energy and the energy content of a rest mass $m$ is actually
\begin{align*}
 E=mc^2.
\end{align*}
In our ordinary experience the velocity $v$ we can deal with, even when flying on a jet, are very much smaller than $c$, so that $v^2/c^2\ll 1$ and the above formula can be approximated by
\begin{align*}
 \frac {L_0-L}{L_0}\approx \frac 12 \frac {v^2}{c^2}, \qquad \frac {\Delta t-\Delta t_0}{\Delta t_0}\approx \frac 12 \frac {v^2}{c^2}.
\end{align*}
To get an idea, for a very fast supersonic jet having a speed $v=3600km/h=1000m/s$ we see that the discrepancy in the measure of lengths or in time intervals is of the order
\begin{align*}
 \frac 12 \frac {v^2}{c^2}\approx 5.5\ 10^{-11}
\end{align*}
that is half of a tenth of billionth! With these numbers it is impossible for us in our daily experience, as well as for Newton and the scientists of its and later epoch, to discover the relativity of time and lengths. This explains why we have the 
{\it wrong} perception that the time is absolute. But things are different in the accelerators or for cosmic particles.\\
Finally, the last formula, $E=mc^2$, is at the basis of nuclear physics and, as all we know, unfortunately, of nuclear weapons. For us it will be particularly interesting for another reason, as we will now see.

\subsection{Einstein's critics to Special Relativity}
Even though it may appear already spectacular, Einstein was not satisfied with his theory for at least two reasons.

\

The first reason was that the theory privileged inertial frames. Substantially, there is no way to establish with exactness if a reference system is inertial or not, neither to prove its existence from a theoretical construction. The reason is that to check
if we are in an inertial system  we need to verify that, in the absence of forces, point like masses move of inertial motion. On the other way, to establish the absence of forces acting on a particle, using Newton's laws, we have to check that it moves
of inertial motion in an inertial frame. Therefore, we cannot separate the concept of inertial frame and the one of absence of forces, with the effect that the {\it principle of inertia}, as formulated by Newton, is not at all operative. The best we can do is to 
{\it assume} we are in an inertial frame, at least at some level of approximation, and to check the consistency of our assumption. This makes difficult to consider such frames as privileged. But there is a further reason for such the difficulty. Assume we are in an
inertial system in total absence of gravity. In a cabin at rest in the far space, a scientist performs his experiments for checking the inertiality of his lab. Now, let assume to switch on a uniform gravitational field, which makes each massive body to fall
with the same acceleration $g$. The people walking on the ground will just feel their weight but will remain inertial just like in absence of gravity. But the scientist will start precipitating with his cabin. All the tools he brought with him will fall exactly in the
same way so that the scientist will see all particle persisting in their {\it inertial motion} if they were before. Indeed, he will not be able to verify he is not really inertial. On the opposite, doing his experiments, he will conclude that he is inertial despite
it will not appear to be so to the inertial observers at rest on the ground. His inertial motions do not appear such to the original inertial system, but his ways to identify himself as inertial, by using Newton's laws, will give him a positive answer. This shows that
there is no reason to privilege the original inertial system with respect to the free falling system.

\

The second reason making Einstein not happy with Special Relativity was related to the gravitational field. Special relativity does not admit transmission of information with a velocity larger than the limit speed. This would lead to paradoxes related to the 
fact that superluminal signals would allow to send information to the past. The theory of Maxwell was perfectly in agreement with the existence of a limit velocity (the speed of light in fact) and the information transmitted by an electromagnetic signal, moving
at a velocity $c$, will arrive at destination with a late proportional to the distance traveled. The Moon main distance from the Earth is of about 360000 Km. The light employs nearly 1.2 seconds to travel it, so that the image we have of the Moon is old of
about 1 second. Similarly, the Sun distance is of about 150 Millions of Km so that the light needs 500 seconds to reach us. The position of the Sun we see is old of about 8 minutes and 20 seconds. Since the Earth rotates of 360 degrees in 24h=86400s,
in 500 seconds it rotates of about 2 degrees. 1 degree is essentially the angle amplitude we see the Moon instead of pointlike. Thus, when we see the Sun, its true position is more to the west, nearly about twice the apparent diameter of
the Moon as seen with naked eyes. More in general, astrophysicists look at stars distant billions of lightyears and that ``now'' are ``switched off'' by a long time. The Maxwell equations take perfectly account of this, they depend on time and predict that we
have to see the position of Moon, Sun or a star with the due late. But the Newton's theory of gravity did not! Its law of gravity states that the force the Sun impresses to the Earth (and vice versa) is proportional to the masses of the two bodies and to
the inverse of the square of the distance between their centres of mass. There is no time in this law: the force on the Earth does not depend on time thus, according to Newton, it doesn't depend on the position in which we see the Sun, but on its actual position
``moved west of two Moons''. This means that the gravitational interaction transmits instantaneously with infinite velocity. Of course this is not compatible with special relativity. For Einstein this was not acceptable at all, a new theory of gravity, allowing the
existence of gravitational waves transmitting the interaction at finite velocity as light does for electromagnetic forces, needed to be developed. 

\subsection{Inertial mass and gravitational mass}
To better appreciate the Einstein's reasoning, we need to discuss a further burden that Newton's theory has been carrying with it by a couple of centuries: {\it the equivalence between inertial and gravitational mass}.\\
The inertial mass $m_i$ of a particle appears in the Newton's law
\begin{align*}
 m_i \vec a= \vec f,
\end{align*}
where $i$ stays for inertial. If we apply a force $\vec f$ to a point particle then its acceleration will be smaller as $m_i$ is larger. It is a measure of inertia, that is of the ``fatigue'' we need to modify the inertial motion of a particle in a given time interval.
If we try to stop a bolt we will need different efforts if it is of polystyrene or of lead. The inertial mass measures the content of matter, a measure different from the volume.  \\
According to the Newton's law of gravity \cite{Newton}, the attractive force also depends on the quantity of matter in the involved particles. Indeed, such law gives us a further way to measure the quantity of matter responsible for the gravitational attraction. Let
us call $Q_1$, $Q_2$ the quantity of matter in two particles distant $d$. Then, their reciprocal attractive force will be
\begin{align*}
 F=\frac {Q_1Q_2}{d^2}.
\end{align*}
If we consider two identical particles, so that $Q_1=Q_2=Q$, we can measure $Q$ by measuring $F$ and $d$ and then getting $Q=d\sqrt F.$ This measure of quantity of matter may be different from both the volume and the inertial mass. It is called the
gravitational mass. But the experiments show that it is exactly equivalent to measuring the inertial mass. If in an inertial frame we fix a particle $Q_1$ and leave a second one, $Q_2$, of inertial mass $m_2$, free to move under the action of the first one, then 
it will acquire an acceleration $a_2$ such that
\begin{align*}
 a_2=\frac {Q_2}{m_2} \frac {Q_1}{d^2}.
\end{align*}
Experiments show that $a_2$ does not depend on the particle $Q_2$, which implies that 
\begin{align*}
 \chi = \frac {Q_2}{m_2}
\end{align*}
is a universal constant independent from the chosen particle. Once we know the value of $\chi$, we also know $Q_i$ any time we measure $m_i$. Using $Q=\chi m$ we can write for the gravitational force
\begin{align*}
 F=\chi^2 \frac {m_1 m_2}{d^2}
\end{align*}
and we get that $\chi^2=G_N$ is the Newton's constant. In this sense, $G_N$ is a universal constant, it has the same universal role of $\chi$. This fact constituted a serious conceptual problem for hundreds of years:
{\it how it may be possible that the same quantity responsible of the laziness of a particle to be accelerated is also responsible to their reciprocal acceleration when two of them are placed not too much far away each other?}
It looks contradictory.

\subsection{Einstein's elevator}
For Einstein this paradox, more than a conclusion, represented a starting point. As characteristic in his methodology, he flips the cards on the table, abandons the preconceived assumptions, and adopts the equivalence between inertial mass
and gravitational mass as a fundamental principle. Starting from this he solves all the problems. \\
Let us go back to the scientist in the cabin. We said that, while falling in a uniform gravitational field, nothing would convince him to be in a non inertial frame. Because of their gravitational masses, he with his test particles will fall with the same acceleration,
appearing inertial to him.
Suppose now that the cabin is really at rest in an inertial frame, whereas the ground where the other observers are placed be the floor of an elevator lifting with acceleration $g$. What changes? Nothing, the description of the system remains 
exactly the same if we simply change the gravitational masses of the scientist and its test particles with the respective inertial masses. The symmetry among the situations is equivalent to the identification between inertial mass and gravitational mass.
The scientist can consider himself as inertial and suppose that the outside peoples are lifting on a large elevator. On the opposite, the external observer can consider themselves as inertial, but in a gravitational field in which the scientist is falling down.
Who is right? Both, since the two descriptions are perfectly equivalent if gravitational mass and inertial mass are the same thing. The only one who is wrong is poor Newton.\\
Therefore, there is a strict relation among gravitational fields and non inertial frames (and, as a consequence, the inertial ones). For Einstein, the uniform gravitational field is perfectly equivalent to the field of apparent forces due to the motion of
a non inertial elevator. For Newton, they were just false forces, but for Einstein have the same rights as a gravitational field. \\
But not all gravitational fields are ``perfectly'' equivalent to a non inertial frame. For example, let us consider the gravitational field on the Earth. It is constant (in time) but not uniform: all forces it exerts are directed toward the center of the Earth.
In this case, if the scientist perform an experiment for a long time, will see that two horizontal particles move one toward the other very slowly (in converging to the center of the Earth), while two vertical particles will move slowly apart (since
the lower one is nearest to the Earth and, therefore, is subject to a slightly larger acceleration). Smaller is the cabin longer is the time interval he needs to verify this phenomenon. It is only in this exact way that inertial systems can exist from the 
Einstein's viewpoint: in presence of a true gravitational field, not eliminable by changing frame as in the uniform case, they exist only locally, approximately, and are defined by free falling bodies. \\
With this vision, for example, the Moon orbiting around the Earth is not accelerated by a gravitational force, but, instead, is just satisfying locally an inertial motion thus following the trajectory of a free motion, the nearest motion to the
to a rectilinear uniform motion. Why the trajectory is then curved? Because space is not flat but curved. This is the effect of gravity in Einstein's theory: {\it space and time are irreparably curved!} \\
Einstein thus discovers that to include gravity in his theory of relativity, he cannot use the flat Minkowski spacetime but a deformed spacetime. Its curvature represents the effect of gravity. If the spacetime is not curved, it is possible to find coordinates
where the gravity completely disappears and spacetime reveals to be Minkowski spacetime. But in general this is not true and the description of our world must be expressed in terms of differential geometry, which we cannot discuss here.

\subsection{Comments on non Euclidean geometry}
Non Euclidean geometry arises essentially when the usual rules of geometry we are used to draw on a flat sheet of paper fail to be true. In two dimensions, this means essentially, for example, that the theorem by Pitagora does not longer applies, that through 
two points may pass more than two ``straight'' lines and so on. We can have a taste of such the geometry by thinking at the Earth surface where we live, assuming it to be a perfect sphere. What is the shortest path to move from a point of the equator,
say, to the north pole? Of course, it is not a straight line! It can be shown that it is the path along the meridian from our starting point to the north pole. And to move between two points on the equator is convenient to stay on the equator. In general,
between two any points on our globe it is convenient to move along the diametral circle passing through these points, as such points belong to an equator. This kind of lines are called {\it geodesics} and mimic the straight lines on a flat plane.
Notice that if as points we take the north and the south poles, then there are infinite geodesics passing through them (all the meridians). Also, we can construct triangles (whose edges are ``straight'' lines) whose sum of internal angles is larger
than 180 degrees: starting from the north pole, move along the 0th meridian toward the equator; then, move along the equator until meeting the 90th meridian and, then, move along the last one toward the north pole. You get an 
equilateral triangle with three right angles. And, of course, it cannot satisfy the Pitagora's theorem. This is the effect of deviation from Euclidean geometry in two dimensions. If we consider very small regions we will have some difficulties to notice
that the sphere is not flat. A small circle of radius $r$ on a flat plane has area $\pi r^2$. But a small space of radius $r\ll R$ on a sphere of radius $R$ has area $\pi r^2 (1+\frac {r^2}{4R^2}+\ldots)$, where the \ldots indicate much smaller corrections.
The area of the ``curved'' circle depends on the radius of the globe via the $1/R^2$ term. This $1/R^2$ thus measures the curvature of the sphere.

\subsection{General Relativity}
In a curved spacetime, very different from Minkowski spacetime, it is difficult to recognise the principle of inertia. Looking at the Moon and the Earth, which move freely along geodesics, if we try to interpret their motion on the basis of the principle
of inertia then we will forced to imagine that they interact with a reciprocal force that, in a very good approximation, is comparable to the Newton's force. But they are just following (and changing) the geometry.\\
We learn from experience that gravitation depends on the disposition of matter in universe, which must be reflected in Einstein's theory as the fact that the geometry of spacetime must depend on how matter moves and vice versa
\cite{Einstein-2} \cite{Einstein-3} \cite{Einstein-4} \cite{Einstein-5} \cite{Einstein-6}. Therefore, one expects a direct relation between geometry of spacetime and matter distribution, which interact reciprocally. Spacetime is no more just a background, 
but a dynamical object. In Newton's gravity the source of gravity is the mass.
But we have seen that relativity shows that the mass is just a form of energy and, therefore, we expect that general relativity should be described by a relation between curvature and energy distribution. This will be the role of Einstein's equations.\\
Even restricting to Minkowski spacetime, we recall that Einstein's theory does not privilege any particular reference frame and, therefore, it admits frames related by totally arbitrary transformation rules. How can we keep control on physics in this
possible chaos of frames? A given formula can change radically form in passing to another frame defined by very complicate transformations.\\
In order to give an answer, let us consider an example introduced by Einstein's in his popular exposition \cite{Einstein-7}. An observer $O$ is at rest in an inertial frame. By hypothesis, there is no gravity there. He imposes to a second clock $O'$ to 
rotate in a circular orbit in a uniform motion (t.i. with constant modulus of the velocity), by means of a thin rope. Keeping fixed the number of laps per minute, longer is the rope, higher is the velocity of $O'$. Of course, the rope is tensioned since it is 
imparting a force to $O'$. $O$ and $O'$ are identical by construction, but, looking at $O'$, $O$ will see $O'$ to tick slower, according to Special Relativity. Now, consider a third clock $O''$, placed in the same place of $O$, but rotating on itself in the 
center of a carousel, rotating in such a way that $O'$ looks at rest with respect to it.
He (we identify the clocks with the observers) can use the same clock of $O$ to measure time (they overlap and, therefore, will agree on that clock), and looking at $O''$ will see the at rest clock to be slowed down, despite it is at rest. 
Furthermore, he will se the rope tensioned (because of centrifugal forces acting on $O'$, would say $O$). We cannot apply Special Relativity to $O''$ since this frame is not inertial. But we can apply General Relativity: according to it, $O''$ will interpret 
the (apparent) centrifugal forces acting on $O'$ as a (real) gravitational field and will conclude that the gravitational field slows down the flow of time. $O$ and $O''$ thus will agree on the comparison of the time flowing as seen by $O$ and by $O'$, 
but not at all on the physical origin of the phenomenon of time dilation: $O$ says it is caused by kinematical reasons, $O''$ says it is caused by gravity. Who is right? The new Einstein's principle states that {\it both are right}. Both explanations are 
equally correct, are equivalent since can be obtained each other by moving from a reference frame to the other one. How can we accept this? The point is that the details of the explanations are not relevant for physics itself, are necessary just for the 
details of the measurements we need to do physics, which, of course, depend on the reference frame (and in general they define it).
The universal fact is that $O'$ ticks slower than $O$ (and $O''$), and every one will agree with this. This is the point: {\it the physical laws have to be written selecting which physical quantities are independent from the detailed explanation.}
Convenient tools for doing it are {\it tensors}, mathematical objects that can be defined in an absolute way, independent from the choice of a reference frame but that assume different descriptions when specified in different frames. We say that 
physical laws must be written in tensorial form, so that they remain valid in any specific frame. \\
Notice that in the ideal experiment we have just considered, the gravitational field can be completely eliminated by choosing a suitable frame (indeed any inertial frame). Nevertheless, the result obtained by applying Einstein's relativity, that is the fact that
gravity affects the flowing of time, must remain true also in presence of not eliminable gravitational fields.

\subsection{The Einstein's equations} We will not write explicitly the quite tricky Einstein's equation, but only symbolically, discussing just as much as needed to appreciate their meaning. We have seen that Einstein's theory of gravity must relate the
distribution of energy and the way it varies in space and time, to the curvature of spacetime. Indeed, Einstein's equations compare two tensors, one describing geometry and the other the mass distribution.
\subsubsection*{Einstein's tensor} Gravity is described by a geometric tensor $\pmb G$, called the {\it Einstein's tensor}. It measures in a very precise way the spacetime curvature. It has the dimension of the inverse of the square of a length,
like we have seen for the two dimensional sphere, compatibly with the fact that curvature is as larger as smaller is the length scale at which deviation from flatness becomes sensible. Despite this, $\pmb G=0$ does not mean that spacetime is flat.
This condition describes the gravitational field in vacuum: a residual curvature survives and it is of fundamental importance since, as we will see, is responsible of the transmission of information between different regions containing matter. If we
imagine spacetime as a sort of a stretched rubber tarpaulin deformed by more or less weighty balls, the vacuum Einstein's equations describe the deformations in points not touching the balls.  
\subsubsection*{Energy-momentum tensor} The distribution of energy is described by a tensor $\pmb T$, called the {\it energy-momentum tensor}, which has the dimension of an energy per unit volume, an energy density. The fact is that in general
one cannot assign energy to a point. If we take an apple (just to evocate Newton) we may know its total energy or the energy in a piece or in smaller and smaller pieces having very tiny volumes. When such volumes go to zero, the corresponding energy
also goes to zero, and what can remain finite is the ratio between the energy and the volume it is contained in. Therefore, energy-momentum tensor says us how much energy is contained in any region as small as we want, and how it changes in space 
and in time.
\subsubsection*{Einstein's equations} The Einstein's equations have the tensorial form
\begin{align*}
 \pmb G=\kappa \pmb T,
\end{align*}
where $\kappa$ is a constant having suitable unit of measure in such a way that, like the l.h.s., also the r.h.s. has the dimensions of the inverse of the square of a length. Indeed, Einstein proved that at low energy densities and at velocities much
smaller than $c$, his equations reproduce to Newton's gravity if
\begin{align*}
 \kappa=8\pi \frac {G_N}{c^4}.
\end{align*}
Recall that in standard units the speed of light is $c=299792458m/s$ and the Newton's constant is $G_N=6.67 \cdot 10^{-11} J\cdot m/Kg^2$, so that the Einstein's constant is $\kappa\approx 2\cdot 10^{-43}m/J$. This is a very small number: 2 divided
by ten 43 times! This means that are necessary enormous energy densities to get appreciable deformations of spacetime. Otherwise, we must look at very large regions to be able to appreciate effects of curvature. This is like to say that spacetime is an
extremely rigid medium, very hard to deform. This is why gravity is so weak. \\
Before passing to the main object of our interest, there is an important point regarding Einstein's equation, which deserves to be mentioned. These equations contain also the equations of motion for matter, since they impose the conservation of energy
and all the conserved quantities related to motion, in a way that is compatible with the equations describing the evolution of the possible other fields (electromagnetic, nuclear, \ldots). Therefore, Einstein's equations have to be imposed together with the
equations of all other present fields, the corresponding initial condition, including for matter whenever it is not microscopically described in term of fields. In most cases, indeed, matter is described macroscopically by thermodynamical quantities,
like energy density $\rho$, pressure $p$ and temperature $T$. These are not independent in general, but related by equations not deducible from Einstein's equations, and which characterise matter in an independent way, so that they must be imposed 
separately from all other equations. These relations, thus independent from the equations of motion, are called {\it state equations}, and, typically, have the form $f(\rho,p,T)=0$. In several situations the temperature is irrelevant, so that the state equation 
takes the form $p=g(\rho)$, t.i. the pressure is a function of the energy density only. We will see later the relevance of the state equation for our aims.

\newpage
\section{Gravitational waves}
We are now ready to discuss the main topic of this short communication. 
\subsection{Prediction of gravitational waves}
One of the most important predictions of Einstein's equations is the existence of gravitational waves. Let us go back to the example of the interaction between the Sun and the Earth. They are moving one around the other just because
are following the geometry of spacetime, which indeed depends on their motion. What Einstein's equations predict is that the deformation of the geometry around the Earth, due to the relative motion of the Sun, does not happens instantaneously,
according to the displacement of the Sun, but it propagates with finite velocity from the Sun to the Earth. It is something like what happens in a lake when we try to move a floating cork stopper by shacking the water with a hand: the stopper will start 
moving only when the waves we generate will reach it. In the same way, the Sun, moving ``inertially'', shacks the space around it and this shaking will be transmitted to the Earth (and vice versa) by waves. These are the gravitational waves. And the Einstein's
equations predict that they move with velocity $c$, that is the speed of light. The propagation of these spacetime deformations in vacuum is governed by the equation $\pmb G=0$. These equations thus say us how the information on what is happening
around the Sun (or everything is generating the waves) are diffused all around in vacuum. Since their speed is exactly the same as for the light in vacuum, when we look at a beautiful sunset we are not only looking at the position of the Sun
eight minutes and twenty seconds ago, but we are also feeling its gravitational attraction as exerted by it from that position eight minutes and twenty seconds ago.\\
This way, Einstein solved all his worries about gravitation and relativity. He predicted the existence of the gravitational waves in 1918 with his paper {\it \"Uber Gravitationswellen} \cite{Einstein-8}, with a bitter taste to have made an intellectual discovery that
never would have the possibility to be detected experimentally due to the incredibly small value of the constant $k$. We recall that $\kappa \approx 2\cdot 10^{-43}m/J$. In the mentioned paper, Einstein computed the evolution of the gravitational waves
showing their extreme weakness and the impossibility to measure their effects on matter. However, there were two issues that were unimaginable at that time and that Einstein did not consider: the existence in the cosmos of astral monsters, like 
black holes and neutron stars, and the very rapid and impressive technological evolution in the short lapse of time of less than a century.

\subsection{Monsters and ogres}
The usual matter we find around is essentially empty. It is made by atoms, whose diameter is of the order of $10^{-8}$cm, 10 billionths of centimetre. It is composed by a nucleus surrounded by minuscule electrons. The nucleus contains most of the
mass, the electrons weighting less than 1/1000 of it. The diameter of the nucleus being nearly 10 millionths than the one of the atom, so we see that matter is concentrated in a very small region. If nuclei were large as an orange the atoms would have
a diameter of few kilometres and we would see a lot of empty space among nuclei. Therefore, the density of matter is much lower than it could be if electromagnetic field (and not only it) was absent.

Now, let us look at a star. It is an enormous mass of gas, mainly Helium and Hydrogen, collapsing on itself because of gravity and its huge mass. When the gas is sufficiently condensed, the molecules start hitting each other with higher frequency
and violence, increasing the thermal agitation. At a given point, when the scattering energy is sufficiently high, nuclei and electrons begin separating. Because of the very high density, due to the gravitational collapse, despite the electric 
repulsion, the protons of Hydrogen begin to interact directly, giving rise to a new physical phenomenon: the {\it nuclear fusion}! The nuclei of Hydrogen, also interacting with electrons, fuse together generating Helium and so producing a lot of energy.
The star is switched on. The energy generated by fusion is so high that the scatterings among the nuclei are able to stop the gravitational collapse. At least until there is Hydrogen to burn. When almost all Hydrogen is burned out the star will start 
collapsing again.
But, then, it is the Helium that starts a fusion process producing heavier nuclei, like Boron and Lithium, and so on. However, at a certain point this process will stop, since it is not possible to generate heavier and heavier nuclei always producing energy.
Starting from Iron and heavier nuclei, fusion becomes endothermic, this meaning that we have to add energy in order to generate fusion. Thus, when there is nothing left to burn the collapse will proceed. However, not indefinitely.  Quantum physics now
enters the game: the Pauli's exclusion principle \cite{Pauli-1} \cite{Pauli-2}, the same that forbids two electrons to overlap, will contrast the gravitational effect with an increasing pressure, which, if the total mass of the star is not too much, will stop the 
whole process. The star will then find a definitive equilibrium, it has been became a white dwarf. ``Not too much large mass'' means of the order of the mass of the Sun. More precisely, the Pauli's pressure can stop the collapse of the star for masses no 
larger than 1.4$M_\odot$, where $M_\odot\sim 2\cdot 10^{30}$kg is the mass of the Sun. This is known as the Chandrasekhar limit \cite{Chandrasekhar-1} \cite{Chandrasekhar-2} \cite{Landau}.  \\
Things change for stars heavier than the Chandrasekhar limit. The Pauli's pressure among electrons (and also among protons) does not suffice to stop the collapse, which continues. Electrons are pushed down inside protons by the enormous
pressure, transforming them in neutrons and generating neutrini. Almost all matter becomes made of neutrons, which are fermions and, again, satisfy the Pauli's exclusion principle. The neutrons are now compressed in a much smaller space than in 
the case of a white dwarf, essentially are ``overlapping'' each other. The Pauli's pressure due to neutrons is thus much larger than the one due to electrons and, if the mass is not too large, it can stop the collapse. In this case the star becomes
a {\it neutron star}, a monster looking as a giant nucleus made of neutrons. We can have an idea of its density: since we have seen that the diameter of a nucleus is 100000 times smaller than the diameters of a an atom, in the volume of an atom
there can belong $100000^3=10^{15}$ nuclei. This means that the mean density of a neutron star is expected to be $10^{15}$ larger than the density of water, $\rho \sim 10^{15}g/cm^3$, 1 billion of tons per cube centimetre! For example, the Earth
has a mean density of $5.5g/cm^3$ for a diameter of nearly $12700km$. In order to get the same density of a neutron star we would need to compress the Earth until reducing it to a diameter of less than $300m$.\\
But, again, we have said that this is the final stage of a collapsing star if the mass is not too large. But what exactly too much large means is not yet known. The new limit value depends on the state equation of the ``nuclear matter'' constituting
the star, which is not known since both theory and experiments are lacking for so high density matter. Observations suggest that the limit mass should be of the order of $2M_\odot$, two solar masses.
For higher masses, even the Pauli's pressure of neutrons becomes insufficient to contrast gravity and, plausibly, no physical process is able to stop the collapse. Therefore, the mass shrinks down below a size determined by a critical radius, known as the
{\it Schwarzschild radius}
\begin{align*}
 r_S=2 \frac {G_N}{c^2} M,
\end{align*}
where $M$ is the collapsing mass. When this happens the stars becomes a new monster, a {\it black hole}. A black hole is one of the most strange objects living in our universe. It is a sort of an ogre, defined by a spherical shield of
radius $r_S$, which can be crossed only in one direction, from outside inward. This way, a black hole devours everything it meets on the road increasing its radius proportionally to the mass. Using the above formulas, the values of $G_N$ and $c$
and recalling that the mass of the Earth is around $5.9\ 10^{24}$kg, we see that if we were able to compress the Earth until he becomes a black hole, its radius would become of about 9mm, roughly like a cherry! Since nothing can escape the 
horizon of a black hole\footnote{at least at classical level. Hawking has shown that, when quantum effects are included, black holes are expected to loose mass emitting a thermal radiation, the {\it Hawking radiation}. However, for black holes
of large masses this effect is very weak and always negligible.} these ogres are essentially more elementary and simpler to be described than neutron stars. We will explain what it means when discussing the experimental results.

\subsection{Measuring gravitational waves}
Gravitational waves propagate with velocity $c$, carrying an energy that deforms time and space in the plane transversal to the direction of propagation.

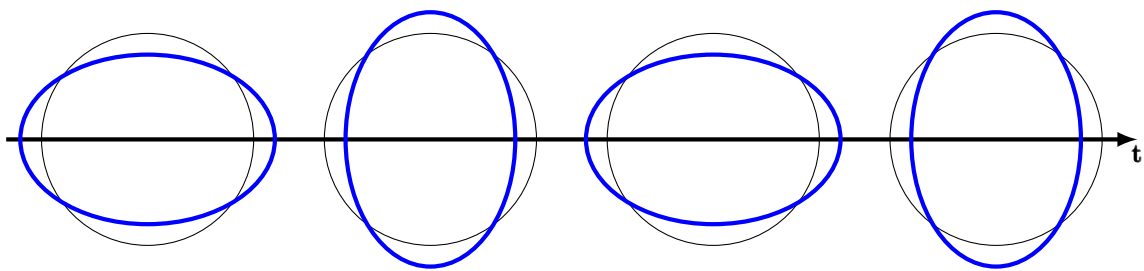
\begin{figure}[!htbp]
\begin{center}
\begin{tikzpicture}[>=latex,scale=0.94]  
\draw [ultra thick,->] (-8,0) -- (8,0);
\draw (-6,0) circle (1.5cm);
\draw [ultra thick, blue](-6,0) circle [x radius=1.8cm, y radius=1.2cm];
\draw (-2,0) circle (1.5cm);
\draw [ultra thick, blue](-2,0) circle [x radius=1.2cm, y radius=1.8cm];
\draw (2,0) circle (1.5cm);
\draw [ultra thick, blue](2,0) circle [x radius=1.8cm, y radius=1.2cm];
\draw (6,0) circle (1.5cm);
\draw [ultra thick, blue](6,0) circle [x radius=1.2cm, y radius=1.8cm];
\node at (8,-0.2) {\pmb t};
\end{tikzpicture}
\end{center}
\caption{A gravitational wave is moving in the direction orthogonal to the plane of a circle. The circle appears to be deformed in different ways at different times.}
\end{figure}

\noindent  Very far from the source, where we are, a gravitational wave can be seen as the superposition of plane waves, each one oscillating with given frequency. 
They deform space in such a way that, when it is contracted in a given direction, it is expanded in the orthogonal direction on the deformation plane. \\
In order to understand better what is happening, let us suppose to have two free masses at rest in a given inertial frame, in absence of gravity, that is in a flat spacetime. Suppose they are placed on the $x$ axes at the positions $x_1=-L/2$
and $x_2=L/2$. At a certain point, they are hit by a gravitational wave moving in the $z$ direction. Spacetime will be deformed but how can we detect it? A way is to consider a laser beam which is shot, say, from particle 1 to particle 2, with a given
wavelength $\lambda$. We can measure the distance between the two masses by counting the wavelengths between $x_1$ and $x_2$. In a flat space it will be $n=L/\lambda$, not necessarily an integer number, of course. Equivalently, we can measure
the distance by measuring the time light employs to travel from $x_1$ to $x_2$: $\Delta t=L/c$. But when the gravitational wave is passing, spacetime is deformed, the running of time changes and, in general, the notion of inertial frame is lost so that
we have not a natural choice of frame where performing measurements. We know that the result of a measure may depend on the choice of a frame, therefore, according to Einstein, we need to look for information that will be independent from the
choice of a specific frame. Now, we know that in presence of a gravitational field one can define local inertial observers. Let us consider such observers placed along the laser beam: they are exactly the observers each one seeing the light of the
laser traveling with velocity $c$ and wavelength $\lambda$. Different observers would not be generically inertial and would measure different values for $\lambda$ and for the speed of light. All together, these observers define a frame (not globally inertial!)
in which it makes sense to define the distance between the two masses as the time $\Delta t'$ employed by light in moving from the first particle to the second one, or, equivalently, the number $n'$ of consecutive wavelengths of the laser between the 
two points. This is called {\it the proper distance} between the two masses. It is clear that, the number of wavelengths is independent on the way we measure it (whereas the traveling time and the measure of $\lambda$ may change with the reference frame)
and it is thus a very convenient way to measure the proper distance. The effect of the transit of a gravitational wave is indeed to modify the proper distance between two free masses by a small amount proportional to the intensity of the wave, usually
called {\it the strain} $h(t)$ of the wave: $\Delta t\to \Delta t'= \Delta t+\delta t$ or, equivalently, $n\to n'=n+\delta n$. $\delta n$ is determined by $\delta t$ which, if the distance between the masses is much smaller than the wavelength of the gravitational
wave, results to be proportional to $h(t)$. Thus, a measure of $\delta n$ provides a measure of $\delta t$ and, then, a measure of the strain $h(t)$.\\
The reason why the distance $L$ between the free masses must be much smaller than the wavelength $\ell$ of the gravitational wave, is that this way the flying time of the laser beam, $L/c$, is sufficiently short so that $h$ is substantially constant
and $\delta t$ depends only on $h(t)$. Indeed, $h(t)$ changes considerably in time intervals in which the gravitational wave travels a displacement of the order $\ell$. Since also gravity propagates with velocity $c$, we get the above condition.
If this condition is not satisfied, the total effect on the proper distance will not be proportional to $h(t)$ but to its average over an interval of time $L/c$. This makes less immediate to have information on the instantaneous value $h(t)$. Nevertheless,
as we will see, it is necessary to consider large proper distances, otherwise the effect of the strain, $\delta L\propto hL$, will be too small to be detectable.\\
Another important observation is what a rigid body is. A rigid body is something made of particles, say, strongly glued each other, by electromagnetic forces. These are non-gravitational forces that, therefore, will react to deformations
of spacetime by accelerating particles with respect to the (local) inertial frames. So, a body is rigid if its proper length is not affected by a deformation of spacetime, thanks to the effect of the
non gravitational forces that hold it together. So, if in place of two free masses we consider two masses hold together by a rigid structure, then, their reciprocal proper distance will not be affected by the transit of a gravitational wave.\\
This are the principles based on which one can design an {\it interferometric gravitational antenna}. The idea is to adapt the Michelson and Morley interferometer, constructed inside a rigid structure, allowing a part of one of the two arms
to be inertially free so that its proper length will be modified by the transit of a gravitational wave. This can be realised by inserting two pendulums along one of the arms, as depicted in the figure below. They are free to oscillate horizontally
so that their proper distance is sensible to spacetime deformations.\\
The pendulum $P_1$ is completely transparent, while $P_2$ is a perfect mirror. In absence of gravitational waves the system is settled in such a way to have in $F$ a disruptive interference so that the screen will appear completely black.  
This means that the path of light along the two arms of the interferometer are such that the split beams meet again in $D$, and then in $F$, with opposite phases. 

\begin{figure}[!htbp]
\begin{center}
\begin{tikzpicture}[>=latex]  
\draw [purple] (-3,0) -- (0,0);
\draw [purple,line width=1pt,opacity=0.8] (-3,0) -- (0,0);
\draw [purple,line width=1.5pt,opacity=0.6] (-3,0) -- (0,0);
\draw [purple,line width=2pt,opacity=0.4] (-3,0) -- (0,0);
\draw [purple,line width=2.5pt,opacity=0.2] (-3,0) -- (0,0);
\draw [purple,line width=3pt,opacity=0.1] (-3,0) -- (0,0);
\draw [purple] (0,0) -- (0,4);
\draw [ultra thick,<->] (-0.5,0) -- (-0.5,4);
\draw [purple,line width=1pt,opacity=0.8] (0,4) -- (0,0);
\draw [purple,line width=1.5pt,opacity=0.6] (0,4) -- (0,0);
\draw [purple,line width=2pt,opacity=0.4] (0,4) -- (0,0);
\draw [purple,line width=2.5pt,opacity=0.2] (0,4) -- (0,0);
\draw [purple,line width=3pt,opacity=0.1] (0,4) -- (0,0);
\draw [purple] (0,0) -- (0.1414,-0.1414);
\draw [purple,line width=1pt,opacity=0.8] (0,0) -- (0.1414,-0.1414);
\draw [purple,line width=1.5pt,opacity=0.6] (0,0) -- (0.1414,-0.1414);
\draw [purple,line width=2pt,opacity=0.4] (0,0) -- (0.1414,-0.1414);
\draw [purple,line width=2.5pt,opacity=0.2] (0,0) -- (0.1414,-0.1414);
\draw [purple,line width=3pt,opacity=0.1] (0,0) -- (0.1414,-0.1414);
\draw [ultra thick,<->] (2.5,-0.6414) -- (8,-0.6414);
\draw [purple] (0.1414,-0.1414) -- (8,-0.1414);
\draw [purple,line width=1pt,opacity=0.8] (0.1414,-0.1414) -- (8,-0.1414);
\draw [purple,line width=1.5pt,opacity=0.6] (0.1414,-0.1414) -- (8,-0.1414);
\draw [purple,line width=2pt,opacity=0.4] (0.1414,-0.1414) -- (8,-0.1414);
\draw [purple,line width=2.5pt,opacity=0.2] (0.1414,-0.1414) -- (8,-0.1414);
\draw [purple,line width=3pt,opacity=0.1] (0.1414,-0.1414) -- (8,-0.1414);
\draw [purple] (0.1414,-0.1414) -- (0.1414,-4.1414);
\draw [purple,line width=1pt,opacity=0.8] (0.1414,-0.1414) -- (0.1414,-4.1414);
\draw [purple,line width=1.5pt,opacity=0.6] (0.1414,-0.1414) -- (0.1414,-4.1414);
\draw [purple,line width=2pt,opacity=0.4] (0.1414,-0.1414) -- (0.1414,-4.1414);
\draw [purple,line width=2.5pt,opacity=0.2] (0.1414,-0.1414) -- (0.1414,-4.1414);
\draw [purple,line width=3pt,opacity=0.1] (0.1414,-0.1414) -- (0.1414,-4.1414);
\shade [left color=cyan!20!white, right color =cyan!5!white, rotate=45, opacity=0.6] (-1,0) -- (1,0) -- (1,-0.2) -- (-1,-0.2) -- cycle;
\draw [ultra thin, rotate=45,opacity=0.6] (-1,0) -- (1,0) -- (1,-0.2) -- (-1,-0.2) -- cycle;
\shade [top color=gray!20!white, bottom color=black] (-1,4) -- (9,4) -- (9,4.2) -- (-1,4.2) -- cycle;
\shade [right color=blue!20!white, left color=blue!70!white, opacity=0.6] (2,-1) -- (2,1) -- (2.5,1) -- (2.5,-1) -- cycle;
\shade [right color=blue!20!white, left color=blue!70!white, opacity=0.6] (8,-1) -- (8,1) -- (8.5,1) -- (8.5,-1) -- cycle;
\filldraw [cyan!10!white] (-4,-4.1414) -- (4,-4.1414) -- (4,-4.2414) -- (-4,-4.2414) -- cycle;
\draw [ultra thin] (-4,-4.1414) -- (4,-4.1414) -- (4,-4.2414) -- (-4,-4.2414) -- cycle;
\draw [thick] (2.25,1) -- (2.25,4);
\draw [thick] (8.25,1) -- (8.25,4); 
\node at (-3.3,0) {\pmb A};
\node at (-0.2,0.2) {\pmb B};
\node at (-0.2,3.8) {\pmb C};
\node at (0.3414,-0.3414) {\pmb D};
\node at (7.8,-0.3414) {\pmb E};
\node at (0.3414,-3.9586) {\pmb F};
\node at (2.25,-1.2) {\pmb {$P_1$}};
\node at (8.25,-1.2) {\pmb {$P_2$}};
\node at (-0.75,2) {$\pmb {d_\perp}$ };
\node at (5.25,-0.95) {$\pmb {L}$ };
\end{tikzpicture}
\caption{Gravitational wave antenna}
\end{center}
\end{figure}
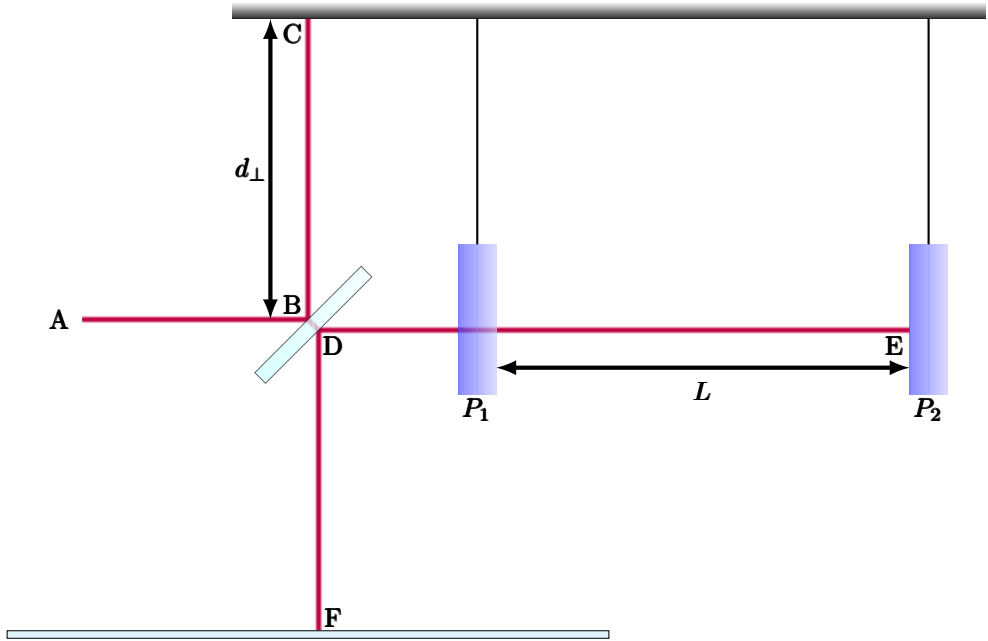

\noindent When a gravitational wave passes in the region of the antenna, the whole structure, being rigid, will not change its proper extensions except $L$, which will change with the proper length, hopefully, accordingly to the Einstein's predictions.
This implies a modification of the length of the corresponding arm, which, even though very small, will no more allow the two beams to meet in $D$ exactly with opposite phases and, therefore, they will not cancel anymore thus projecting on
the screen a visible interference picture. This picture and its intensity depends on the $\delta n$ generated by the deformation, so we get a measure of $\delta t$ and, thus, of $h$.  
\subsection{LIGOs and VIRGO}
There are three interferometric gravitational antennas in the LIGOs VIRGO collaboration: two LIGOs \cite{LIGO} \cite{Advanced-LIGO} (Laser Interferometric Gravitational-Wave Observatory) in the USA, and Virgo \cite{Virgo} in Italy. One of the two 
labs of LIGO is placed in Hanford Site, in place of an ex nuclear implant, in the state of Washington, near the town of Richland. The other one is in Livingston, Louisiana. They are separated by a distance 
of 3000km, corresponding to about 10ms light, since 10 milliseconds is nearly the time light needs to travel through 3000 kilometres. Moreover, they are equipped each one with a gravitational wave interferometer having an arm of length 
$L\approx 4km$ (between the pendulums). In order to amplify the signal, the laser beam is reflected 280 times between the two pendulums before being redirected to the interference point (the point $D$ in the picture). This way the arm has an effective 
length of 1120 kilometres. \\
They are designed to reveal gravitational waves in a range of frequencies $\nu$ between 30 and 300Hz. Since the wavelength $\lambda$ is related to the frequency $\nu$ by the relation
\begin{align*}
 \lambda=\frac c\nu,
\end{align*}
we see that it varies in the range $1000-10000km$. Therefore, the effective length of the arm is not much smaller than $\lambda$ and accurate formulas, avoiding the oversimplifications we discussed above, are needed in order to correctly
estimate the strain $h(t)$. Moreover, they are expected to reveal signals of the order $10^{-21}$, which means that the proper distance $L$ between the pendulums will vary by a displacement 
\begin{align*}
 \delta L\approx hL\approx 4 \cdot 10^{-21} km= 4\cdot 10^{-16} cm, 
\end{align*}
which is approximately one 200th the diameter of a proton! The pendulums have masses of about $40kg$. Probably a sneeze would generate a displacement of them larger than several atoms. Thus, we understand how much
accuracy must be considered in order to protect the antenna from other causes of oscillations: seismic movements, fluctuations of humidity, temperature and density of air and its flow, noise due to the passage of vehicles or the walking of
scientists near the antenna, variation of the temperature of the components of the antenna (generating dilatations an deformations), thermal, statistical and quantum fluctuations of the characteristic of the laser beam, and so on.  \\
Each of these phenomena can generate very small displacements that, despite their smallness, can be larger than fractions of a proton diameter. One has then to be sure to eliminate or make them clearly distinguishable from the gravitational
effects in order to have an efficient antenna. There is no place here to discuss how, indeed, all this problems are solved, but this shows us the enormous difficulties one must tackle in order to reveal gravitational waves even when the latter are generated
by catastrophic events. And we can understand the incredulity on the part of Einstein for experimentalists to be able to detect such a weak phenomenon.\\
There is a third antenna, Virgo, sited in Cascina, near Pisa, in Italy, built in an Italian-French collaboration. Its interferometer has an arm of $3km$ with 30 reflections, for an effective length of about $90km$. The corresponding amplification
of the signal is much smaller than the one of the LIGO interferometers. In the advanced evolution the effective length will be enlarged to nearly $800km$. The importance to have a third antenna is to be able to triangulate the position of the source
of the signal.

\subsection{The observations}
Up to the date of the present communication, March 1, 2018, there have been detected six events, named GW150914 \cite{GW150914}, GW151226 \cite{GW151226}, GW170104 \cite{GW170104}, GW170608 \cite{GW170608}, GW170814 \cite{GW170814},
GW170817 \cite{GW170817}. GW stays for gravitational wave and the numbers arethe date of the event, GWaammdd. The first five events have been identified as generated by the coalescence of two black holes, while the last one as corresponding to 
the coalescence of two neutron stars. We will discuss the first and the last detections.
\subsubsection*{GW150914} In September 14, 2015, Virgo was not yet active and the LIGO labs detected a signal corresponding to a displacement of the pendulums of about one 200th of a proton diameter, with a starting frequency
of about $35Hz$ increasing fast up to $250Hz$. The whole signal had a duration of about one 5th of second, ending up in a quick ring down. Comparison with theoretical prediction allowed the scientists to understand that this signal 
has been generated by the interaction between two black holes, having masses $M_1\approx 36M_\odot$ and $M_2\approx 29M_\odot$, where $M_\odot\approx 2\cdot 10^{30}kg$ is the mass of the Sun. Like the Moon and the Earth,
they started rotating one around the other with increasing velocity while emitting gravitational waves of increasing frequency (the frequency of the waves is twice the frequency of rotation of the black holes). This way, they gradually lost energy
thus decreasing their reciprocal distance and increasing the rotation energy. The corresponding increasing of the frequency of the emitted gravitational waves is said the {\it chirp} and depends in a very precise way from the masses of the orbiting
bodies. This process happened faster and faster, until the two black holes touched devouring each other and reducing to a final stationary black hole. This final phase corresponds to the ring down. The mass of the final black hole results to be 
about $M\approx 62M_\odot$. There is a loss $\Delta M=M_1+M_2-M\approx 3M_\odot$: three solar masses has been transformed in pure energy, mainly distribute in gravitational waves!\\
This means that the emitted energy is about
\begin{align*}
 \Delta E=\Delta M c^2=5.4\cdot 10^{47}J.
\end{align*}
It is an incredible amount of energy. If 7billions of people would have at disposition 300kw of electric power (100 times the power we have at disposition in Italy) and each of them used the full power uninterruptedly night and day, the energy $\Delta E$
would be consumed in $8\cdot 10^{24}y$, which means 8 millions of billions of billion of years, nearly 600000 billions times the age of the universe! This amazing energy has been exploded in all directions in a fraction of seconds. \\
Why it did not ravaged us? Because it happened at a huge distance, $D\approx 410Mpc$, where $Mpc$ stays for megaparsec, 1 million of parsec. One parsec consists in $3.26$ light years (ly) and 1ly corresponds nearly to
9500 billions of km. Therefore, $D\approx 12.7 \cdot 10^{24}m$. When reaching the Earth, the energy $\Delta E$ is distributed on a sphere of radius $D$ and area $A=4\pi D^2$. Of this, only the portion of area $\pi R^2$ hits the Earth, if 
$R\approx 6366km$ is the radius of the Earth. It is thus a energy of about
\begin{align*}
 \varepsilon =\frac {R^2}{4D^2} \Delta E \approx 34kJ,
\end{align*}
of which only a small fraction (less than 1/10000 of Joule) hits the regions covered by the antennas, leading to a deformation of about a fraction $10^{-21}$. This is exactly what has been measured by the antennas, in perfect agreement
with the predictions of Einstein's theory. Before discussing the meaning of these results, let us describe the other case.
\subsubsection*{GW170817} In August 17, 2017, it has been detected the sixth event, but the first case of gravitational waves generated not by the coalescence of two black holes, but of two neutron stars. This time also Virgo was active, even
if not yet sensitive enough to resolve the signal, this has been sufficient for the triangulation of the position of the source. The observed signal has qualitatively the same form as for the previous cases, but with different qualitative details.
Indeed, it corresponded to the merging of masses below $2M_\odot$, thus compatible with neutron star masses. The frequency of the signal again raised from 40 to 300Hz, then suddenly disappearing in a ring down. The process taken nearly
30 seconds. This collapse of neutron stars, merging probably in a final black hole, happened at 40Mpc, nearly one tenth the distance of GW150914. The energy emitted has been lower than the previous case, but it is estimated to be larger 
than $\frac 1{40} M_\odot c^2$, say of the order 1/100 the one in GW150914. Why then this event is even more exciting than the previous five ones? Because, as we said, it corresponds to the merging of neutron stars in place of black holes.
Despite their large masses, black holes are essentially Einstein vacuum solutions, and, since they devour everything, even when fusing one into the other, nothing can essentially escape but the gravitational waves they generate. \\ 
For neutron stars, the story is quite different, since they are made of matter and this influences the whole process in several ways. First, before they touch each other, the evolution depends also on how matter reacts to the stress of gravity. Indeed, 
especially in the last moments of their ``dance'', when they are very near before touching, they will exert enormous tidal forces, similarly like the Moon does with the Earth, but with an incredible strength. How matter supports such a stress
depends on the state equation describing high density nuclear matter. As we already said, such an equation is not yet known at a theoretical level, and cannot determined experimentally since such huge nuclear densities cannot be
realised in any laboratory. Moreover, when the stars touch, a tremendous bump ejects a lot of neutronic matter everywhere around. The direct interaction of matter also generates electromagnetic waves and neutrinos. For example, the expelled
neutronic matter will appear initially in form of nuclei containing an excess of neutrons. When no more constrained by gravity, neutron are no more stable particles and tend to decay in protons, emitting an electron and an antineutrino. When a
suitable number of neutrons are transformed, that is when they become stable nuclei, typically gold and similar nuclei, this process stops. In any case, it means that beyond gravitational waves, one also will reveal electromagnetic signals, neutrini,
and, therefore, a lot of different information. For this reason, events of this kind are called {\it multi-messengers}.

\subsection{Conclusions and new perspectives for Physics}
GW150914 has been the first direct revelation of gravitational waves, but it has not been the first proof of their existence. In 1974, Russell Alan Hulse and Joseph Hooton Taylor Jr. observed the orbit of a pulsar around a neutron star \cite{Hulse-Taylor}. 
General relativity predicts that two stars orbiting one around the other emit gravitational waves, thus loosing energy and, then, changing their period. This variation of the period can be computed with high precision, but it is very difficult to be measured on 
usual stars because the weakness of the effect. The lost of gravitational energy is higher for a pair of neutron stars, but tracing the orbit of a neutron star is quite difficult since their brightness is very weak. But a pulsar is a neutron star that is rotating very 
fast around its own axis (with ``days'' of the order of few milliseconds) so that, carrying a magnetic field, they generate X-ray pulses with a well defined frequencies and that can be observed from the Earth. If the pulsar is orbiting around a black hole or another 
neutron star, then the pulse is modulated by the orbital motion. This allows to measure the period of the orbit with high precision. Hulse and Taylor used exactly this mechanism to compare the variation of the period per year with what predicted
by Einstein's equations because of gravitational waves emission. They found a perfect agreement, this way obtaining an indirect proof of the existence of gravitational waves. And for this they won the Nobel prize. However, it has not been
(and it is not) possible to detect on Earth such waves because are too weak. \\
However, GW150914 is not only this: it represents also the first direct confirmation of the existence of black holes of the dimensions described above. Notice that the Schwarzschild radii of the two black holes involved in the bump are of the order
of 100km. Recall that nothing can escape from the black hole, not even light. Therefore, we cannot see black holes, usually we can see their effects on the matter around them. This means that we will see nothing if the black holes are not surrounded
by matter. But {\it gravitational waves allow us to see them!} Indeed, it is exactly what happened at LIGO labs. The form of the measured signal gives information both on the masses and the period and one can then follow the change of the orbit
radius until they touch (recall that the third Kepler law shows how the radius of the orbit changes with the period). Therefore, looking at the signal we can see how much the distance between the two bodies can decrease before they touch. If such
the distance is smaller than the sum of their Schwarzschild radii, they cannot be nothing else than two black holes.\\
A third thing is that this kind of events provides new tests for general relativity. Indeed, it is not a priori guaranteed that Einstein's equations must remain true at strong gravities, like in regions where such ogres and cannibal monster are present.
But up to now none of the observed events has required modifications of the Einstein's equations in order to get an explanation. Thus we can consider them as proves of the robustness of general relativity.\\ 
This is true also for GW170817. For example, in this case it happened that the light generated by the terrible bump between the neutron stars reached us 1.7 seconds later, proving that gravitational waves and light move at the same speed, as 
predicted by general relativity. Notice that we are speaking about 1.7s compared to a distance of about 100 millions of light years. Moreover, we have to notice that gravity interacts very weakly so that it travel through cosmological distances
almost without being deformed by the matter it meets along the path. This can account the small delay of light.\\
But there is further information we can infer from the multi-messenger. We said that in the final stage of the dance of the stars, the exact evolution and the final collapse will depend on the relative tidal forces and on the equation of state
governing neutronic matter at incredibly high density and low temperature. Notice that the temperature of a cooled neutron star is of about $10^{10}K$, that is 10 billions of Kelvin degrees. This is nearly one thousand times the temperature at the center of
the Sun. But for a for a neutron star it must be considered essentially as to be zero temperature for the following quantum mechanical reason. Neutrons (and quarks), constituting the neutron star, are fermions, which mean that they have to 
satisfy the Pauli's exclusion principle. As such, they cannot all fall into the minimal energy state after it has been already saturated by other fermions. Therefore, compatibly with that principle, they can just occupy not the lowest energy state but the 
lower energy allowed. At the end they will occupy all the states up to a maximal energy, called the Fermi's energy. This happens truly at zero temperature, or, more practically, at temperatures $T$ such that $kT$, where $k$ is the Boltzmann 
constant, is much smaller than the Fermi's energy. The Fermi's energy depends on the density of neutronic matter and on the equation of state. The latter is not known in our case, but a rough estimation can be made equally, ensuring
that the Fermi's energy for a neutron star, with its enormous density, is much larger than $kT$ for $T$ of the order of ten billion of Kelvin degrees. Since there is no possibility to reproduce these conditions in a laboratory, we have not direct
information on the equation of state of dense cold matter. But we can hope to infer new information on it by reading the details of the final stage of the merging as shown by the gravitational wave signal, as well as the from the electromagnetic
signals. This way, looking at far and giant multi-messengers we can get information on very high energy nuclear physics.\\
In conclusion, measuring gravitational waves opens a new exciting era. It gives us new eyes to look both at the deep and far universe as well as the microscopic world of nuclear physics.

\section*{Acknowledgments}
I would like to thank Carlo Pagani for invitation to give a presentation to the Accademia of Brera. I also thank Simone Noja for helpful comments.

\newpage
\begin{appendix}
\section{An elementary deduction of the Lorentz transformations}
Here we want to show in what sense the principle of inertia, together with some simple assumptions, determines the Lorentz transformations. 
\subsection{Inertial frames}
For sake of simplicity we consider a one dimensional system, so that an inertial frame is described by a unique
spatial coordinate $x$ and a measure of time $t$, so that we will call it $K=(x,t)$. In order to be precise we have to spend some words: \\
Since with inertial motions we mean straight-line motions with constant velocity, the space must be a straight line.
The coordinate $x$ is thus defined by choosing a point on the line, called {\it origin}, and a unit of length to measure distances. The origin separates the straight line into two parts, called positive and negative, and the coordinate of a point is given
by its distance from the origin and with a positive or negative sign accordingly if the point belongs to the positive or negative part of the straight line.\\
The time coordinate $t$ is defined by considering an infinite set of clocks, each one at rest in a fixed point $x$ of the reference frame. All the clocks are then synchronised all together in an arbitrary way. This defines the time $t$ in $K$.\\
It is important to point out that the way to construct the yardstick, the clocks and the way to synchronise them, must respect the physical rules retained to be valid in an inertial frame. Indeed, in a theory that privileges inertial frames, the latter
are all equivalent and, therefore, we must ensure that in all such frames there is the way to construct the reference frames exactly in the same way, with instruments constructed in the same way, and the same procedure of synchronisation.
\subsection{Homogeneity, isotropy and reciprocity.}
Now we make some simple considerations going beyond the principle of inertia. An inertial frame can be characterised by a system of coordinates $(x,t)$ so that, indeed, it individuates a spacetime, not only a space. We want to assume simple properties
of such a spacetime. The simplest property we can require to space time is {\it to be homogeneous}. This means that all points of the straight line are equivalent and all instants of time are equivalent. Therefore, any observer in any give point
on the line cannot recognise where it is and will have the same perception of space if put in the origin or in any other given point. And this perception will not change in time. In particular, this implies that the flowing of time is the same in any point and the clocks
remain synchronised forever. \\
The homogeneity, expressed in this way, also requires {\it isotropy}: the observer in $O$, or in any other point, looking around will see equivalent points and, therefore, the space will appear to him equal in all directions. Of course, on the line there are just two 
directions, forward and backward. In particular, this means that the negative and positive part of the line are equivalent, and fixing which one is positive is just a matter of choice, not implied by any physical imposition. The same is true for the choice of
the origin.\\
There is also another requirement to add to these assumptions, which will be crucial in comparing two distinct inertial frames: {\it reciprocity}. Consider a second inertial system $K'=(x',t')$ with origin $O'$. Since both frames are inertial, any inertial motion in 
each one of them must appear to be inertial in the other. In particular, since $O'$ is at rest in $K'$, it will appear to move with constant velocity by $K$, say with velocity $v$. Viceversa, $K'$ will see $O$ moving with constant velocity, call it $v'$. By isotropy
we can choose the orientation of each frame as we want: let us choose the opposite ones so that $O$ and $O'$ result to be front to front (or back to back). Because of homogeneity and isotropy, the situation is perfectly symmetric and what $O$ sees
looking at $O'$ must be exactly the same seen by $O'$ looking at $O$. In particular, we are expected to conclude that $v=v'$. If we change the orientation of $K'$ so that both frames are oriented in the same way, then we get the change $v'=-v$. 
This is the reciprocity principle: if the two frames are oriented in the same way, if $O$ sees $O'$ moving at velocity $v$ than $O'$ sees $O$ moving at velocity $-v$. Of course the same is true for any other pair of points.
\subsection{Relations between inertial frames.}
The considerations just done do not allow us to get trivial conclusions about the relation between $K$ and $K'$, that is between $(x,t)$ and $(x',t')$. The fact that the physical laws are formulated in an arbitrary inertial frame implies that if we do an 
experiment in $K$ with given possible results, repeating the same experiment in $K'$ we will get the same possible results. But nothing says that if $K$ tries to understand what is happening in $K'$, it will see exactly the same things seen by $K'$ itself.
Consider a point $p$ in $K'$ in the positive part with respect to $O'$. Its coordinate in $K'$ will be $x'$ equal to the length of the segment $O'p$ as measured by $K'$. $K$ sees the segment moving and it has to use a different procedure than $K'$ to measure
its length. So we cannot assume a priory that it will get the same result for its length: the most general assumption is that for $K$ the length of that segment will result to be proportional to $x'$, say 
\begin{align*}
L=b x', 
\end{align*}
where the proportionality factor $b$ is a priori a function of $v$ but nothing else (as follows easily from the symmetry considerations). Also, since a change of orientation changes the sign of $v$, by isotropy it follows that $b(-v)=b(v)$, that is it is an even
function of $v$.\\
Now, let us assume for simplicity that, at $t=0$, $O$ and $O'$ coincide and $O'$ indicates the same instant $t'=0$. Since $O'$ moves with constant velocity $v$, at a time $t$ it will have in $K$ the position $vt$. Therefore, $K$ will assign to $p$, at rest in $K'$
at the position $x'$ as above, the coordinate $x=vt+L$, that is
\begin{align*}
x=vt+b(v) x'. 
\end{align*}
this relates the coordinates $(x,t)$ to $x'$ and is not sufficient to relate the two frames. Nevertheless, $K'$ can repeat the symmetric procedure to relate $(x',t')$ to $x$, with the caveat to exchange $v$ with $-v$. Using that $b$ is even it will get (with the same
$b$ by symmetry!):
\begin{align*}
x'=-vt'+b(v) x. 
\end{align*}
From both relations one easily gets
\begin{align*}
 x&=\frac {x'+vt'}{b(v)}, \\
 t&=\frac {t'}{b(v)}+\frac {x'}{b(v)v} (1-b^2(v)),
\end{align*}
and the inverse (obtained exchanging $(x,t,v)\leftrightarrow (x',t',-v)$)
\begin{align*}
 x'&=\frac {x-vt}{b(v)}, \\
 t'&=\frac {t}{b(v)}-\frac {x}{b(v)v} (1-b^2(v)).
\end{align*}
How can we determine the function $b(v)$? We can consider further composition with a third inertial motion. Let us assume that $K'$ is looking at a particle starting from $O'$ at $t'=0$, with constant velocity $w'$ with respect to $K'$. Therefore,
its coordinate at time $t'$ will be $x'=w' t'.$ On the other hand, the particle moves inertially w.r.t. $K$ also, with a velocity $w$, such that $x=wt$, since at $t=0$ we have $t'=0$ and $O'=O$. Using the above relations expressing $(x,t)$ in terms of
$(x',t')$ we get
\begin{align*}
 w=\frac xt=\frac {x'+vt'}{t'+\frac {x'}{v} (1-b^2(v))}=\frac {w'+v}{1+\frac {w'}{v} (1-b^2(v))},
\end{align*}
where we used $x'=w't'$. We see that the velocity is not just the sum of the velocities! This is the {relativistic composition law} of velocities. \\
Now, consider a third inertial frame $K''$, with origine coincident with the particle just considered. By reciprocity, it will see $O$ and $O'$ moving back with velocities $-w$ and $-w'$ respectively, while $O$ appears as a particle moving at velocity $-v$
with respect to $K'$. Therefore, $K''$ can use exactly the same rule as above to relate the velocity $-w$ he measures for $O$ to the velocity $-v$ measured by $K'$ for $O$, the relative velocity of $K'$ w.r.t. $K''$ being $-w'$ (in place of $v$). Therefore,
$K''$ will get
\begin{align*}
 -w=\frac {-v-w'}{1+\frac {-v}{-w'} (1-b^2(-w'))},
\end{align*}
that is
\begin{align*}
 w=\frac {v+w'}{1+\frac {v}{w'} (1-b^2(w'))}.
\end{align*}
The two expressions of $w$ (as obtained by $K$ and by $K''$) must coincide for any given choice of $v$ and $w'$. This happens if and only if
\begin{align*}
 \frac {w'}{v} (1-b^2(v))=\frac {v}{w'} (1-b^2(w')),
\end{align*}
that is
\begin{align*}
 \frac {1}{v^2} (1-b^2(v))=\frac {1}{w^{\prime2}} (1-b^2(w')).
\end{align*}
Since the right hand side does not depend on $v$, the same must be true for the left hand side, so that it must be a constant with the dimension of the inverse square of a velocity. Let us call it $k=\varepsilon/V^2$, where $V$ is a constant with the dimension
of a velocity, an $\varepsilon=0,1,-1$ according if the constant $k$ is zero, positive or negative respectively. Therefore
\begin{align*}
 \frac {1}{v^2} (1-b^2(v))=\frac {\varepsilon}{V^2},
\end{align*}
which gives
\begin{align*}
 b(v)=\sqrt{1-\varepsilon \frac {v^2}{V^2}},
\end{align*}
where we have chosen the positive root, taking into account that the axes have the same orientation.
For the relativistic transformations and the composition law we finally get
\begin{align*}
 x&=\frac {x'+vt'}{\sqrt{1-\varepsilon \frac {v^2}{V^2}}}, \\
 t&=\frac {t'+\varepsilon \frac v{V^2} x'}{\sqrt{1-\varepsilon \frac {v^2}{V^2}}}, \\
 w&=\frac {w'+v}{1+\varepsilon \frac {vw'}{V^2}}.
\end{align*}
Notice that for $\varepsilon=0$ we get the usual Galilei transformations, with $t=t'$. This can be also obtained in the limit $V\to\infty$. Notice also that for given $V$ and $\varepsilon<0$ any value of $v$ is allowed, while for $\varepsilon>0$ only velocities
lower than $V$ are allowed. However, we see that negative $\varepsilon$ can be easily discarded for the following reason: \\
for given $V$ and positive $w'$ for the particle, we get that if $w'$ is fast enough the sum of velocities can diverge or even change sign. In particular, particles moving forward fast enough with respect to $K'$ would move backward w.r.t. $K$. If we want to 
avoid the singularity, for given (positive) $v$ we should allow $w'$ to vary from $-\infty$ up to $V^2/v$. But this would break isotropy and homogeneity of course.\\
The situation is different for $\varepsilon>0$. What is the meaning of $V$? Of course it says that $-V<v<V$ so it is a limit velocity for $K'$. But being a constant, it must be universal in the sense that it must be the same for any reference frame. This means
that, in our above construction, it must be a limit velocity for $w'$ also. Indeed, this is what happens without giving rise to any contradiction, since if $w'=V$ in $K'$, then, from the composition law, $K$ will measure a velocity
\begin{align*}
w=\frac {v+V}{1+\frac vV}=V, 
\end{align*}
so $V$ is invariant under composition with velocities and is thus the same as seen by any inertial frame!\\
In conclusion, the most general transformation we get naturally, without imposing arbitrary choices, is
\begin{align*}
 x&=\frac {x'+vt'}{\sqrt{1- \frac {v^2}{V^2}}}, \\
 t&=\frac {t'+\varepsilon \frac v{V^2} x'}{\sqrt{1-\frac {v^2}{V^2}}}, \\
 w&=\frac {w'+v}{1+\frac {vw'}{V^2}}.
\end{align*}
We can get the Galilei transformations, with an absolute time $t=t'$ by assuming the assolute velocity to be infinitely large. But it would be just an arbitrary choice, it is Nature that has to say us if, compatibly with inertiality, it does exist a finite 
absolute velocity or not. And, as we discussed in the text, the Michelson and Morley experiment shows that the absolute velocity is not infinite by coincide with the speed of light, $V=c$. \\
It is worth to mention that we have not used this fact in our deduction so, in this sense, the existence of an absolute velocity has nothing to do with the speed of light a priory, but just with the structure of spacetime (inertial properties and homogeneity).
Therefore, the fact that $V=c$ cannot be read as the fact that the relativity transformations have something to do with the light, but rather that, if light signals move at the absolute velocity this must be due to a deep relation between the behaviour of
the electromagnetic field and the properties of spacetime. On the opposite, following this reasoning, we could say that the well tested validity of Maxwell equations, combining the fact that they take the same form in any inertial frame only if the speed of
light in vacuum is $V$ with the principle that all inertial frames should be equivalent, we can conclude that Maxwell equations tell us a way of determine the value of $V$: by measuring $c$.

\section{Contraction of lengths, dilation of time, relativity of simultaneity.}
Employing the transformations of the previous section it is easy to get some consequences about comparison of measurements in two different inertial frames, say $K$ and $K'$ as above. We already know that a segment at rest in $K'$ of length $L'$
will appear to $K$ to have length $L=b(v)L'$, that is
\begin{align*}
 L=\sqrt{1-\frac {v^2}{c^2}}\ L'.
\end{align*}
It thus appears to be contracted from the point of view of $K$. It is clear that the same effect appears from the point of view of $K'$ looking at a segment at rest in $K$.\\
Similarly, suppose $K$ is looking the clock sited in a fixed point in $K'$, say at coordinate $x'_0$. According to $K$ it will correspond to a time
\begin{align*}
 t=\frac {t'+x'_0 v/c^2}{\sqrt{1-\frac {v^2}{c^2}}}.
\end{align*}
The interval of time between two events in $x'_0$ is $\Delta t'=t'_2-t'_1$. Using the above formula and subtracting the corresponding times $t_2$ and $t_1$ we get that the interval as measured by $K$ is
\begin{align*}
\Delta t=\frac {\Delta t'}{\sqrt{1-\frac {v^2}{c^2}}}. 
\end{align*}
A time interval between events in $K' $ will appear to $K$ as expanded by the relativistic factor $1/\sqrt{1-\frac {v^2}{c^2}}$. This does not means that the time flows more quickly in one or the other frame. If $K'$ measures the time interval
between events in $K$ he will get exactly the symmetric result! The only conclusion that we can affirm is that the flow of time is relative, exactly as space, and that in order to measure it we have to specify the events defining the interval of time and the
reference frame in which the measure is performed. Nothing more than this.\\
The above formula also shows that two simultaneous events in $K'$, at different points, are not simultaneous for $K$ (and viceversa). If they happen in the same time $t'$ at the points $x'_1$ and $x'_2$, then they will appear to $K$ to happen
at two different instants
\begin{align*}
t_i= \frac {t'+x'_i v/c^2}{\sqrt{1-\frac {v^2}{c^2}}},\qquad\ i=1,2.
\end{align*}
Since both systems have synchronised correctly each one its own clocks, both are right and the events {\it are} simultaneous in $K'$ and {\it are not} simultaneous in $K$. This is relativity of simultaneity.  \\
All this effects become negligible for velocities $v\ll c$, as it happens in our daily experience.

\end{appendix}


\end{document}